\documentclass{svjour2}                    
\smartqed  
\usepackage{graphicx}
\journalname{J Low Temp Phys}
\begin{document}

\title{On The Mobile Behavior of Solid $^4$He at High Temperatures}



\author{A. Eyal         \and
        E. Polturak 
}


\institute{A. Eyal \and E. Polturak \at
              Department of Physics, Technion, Haifa 32000, Israel \\
              Tel.: +972-4-8292027\\
              Fax: +972-4-8292027\\
              \email{satanan@tx.technion.ac.il}             \\
           \and
           E. Polturak \at
              \email{emilp@physics.technion.ac.il}
}

\date{Received: date / Accepted: date}


\maketitle

\begin{abstract}
We report studies of solid helium contained inside a torsional
oscillator, at temperatures between 1.07K and 1.87K. We grew single
crystals inside the oscillator using commercially pure $^4$He and
$^3$He-$^4$He mixtures containing 100 ppm $^3$He. Crystals were
grown at constant temperature and pressure on the melting curve.
At the end of the growth, the crystals were disordered, following which they partially decoupled from the oscillator. The fraction of the decoupled He mass was temperature and velocity dependent. Around 1K, the decoupled mass fraction for crystals grown from the mixture reached a limiting value of around
35\%. In the case of crystals grown using commercially pure $^4$He
at temperatures below 1.3K, this fraction was much smaller. This
difference could possibly be associated with the roughening transition at the solid-liquid interface.

\keywords{Solid Helium \and Supersolids \and Disorder}
\end{abstract}


    \section{Introduction} \label{sec:intro}

Solid Helium has been intensely studied during recent years as
part of the search for supersolidity \cite{Andreev,Leggett1970}.
These studies revealed an anomaly in the acoustic properties of
the solid around 0.2K \cite{Goodkind1997}, and subsequently a Non
Classical Rotation Inertia (NCRI), first discovered by Kim and
Chan \cite{KC2004} using a torsional oscillator (TO). The TO was suggested as
a tool to search for supersolidity by Leggett \cite{Leggett1970}.
The initial discoveries of the NCRI were later
confirmed and expanded upon by many others
\cite{Shirahama,Kubota,Reppy2007,Kojima,Davis,Kim}. In addition, an increase of the shear modulus \cite{Beamish} and recently mass flow in the solid
\cite{Hallock} in the same temperature range were reported. There
is an ongoing debate regarding the interpretation of these TO and
shear modulus experiments in terms of supersolidity
\cite{Davis,Balatsky,Kuklov,Reppy2010}. We have essentially
repeated the TO experiments with two important differences: (a) we
used single crystals instead of polycrystals used by a majority of
other groups, and (b) we searched for effects similar to those
reported (NCRI) at higher temperatures. We found \cite{Eyal2010} that disordering
a single crystal produces qualitatively the same results observed
at low temperature TO experiments. We have now extended our
temperature range enabling us to grow single crystals between
about 1K and up to nearly 2K. In addition, we did the experiments
using not only commercially pure $^4$He (around 0.3 ppm of $^3$He), but also a mixture
containing 100ppm of $^3$He. Overall, we see velocity dependent
mass decoupling in all our experiments. The results with pure
$^4$He however differ in some respects from those obtained using a
$^3$He-$^4$He mixture. These new results are reported in the
present manuscript.


    \section{Experimental System} \label{sec:experimentalSystem}

\begin{figure}
  \includegraphics[width=0.75\textwidth]{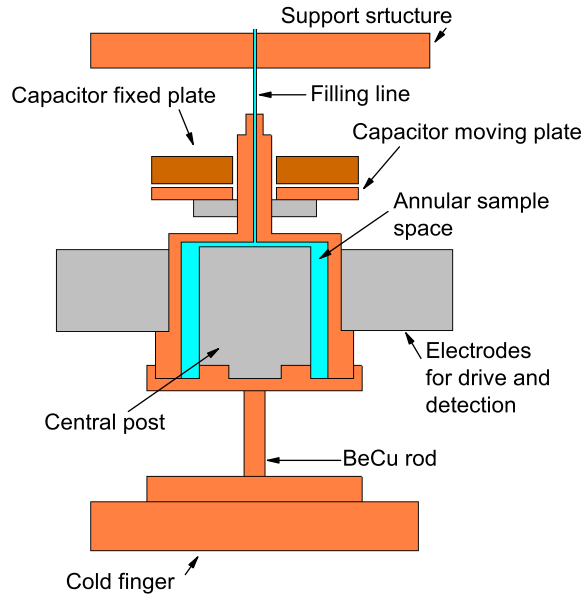}
\caption{(Color on-line) A schematic cross section of the torsional oscillator
(TO) including a capacitance pressure gauge for in-situ pressure
measurements. The drawing is to scale. Crystals are grown inside
an annular sample space. The inner radius of the annulus is 6.5mm,
the height is 10 mm, and its width is 2mm. The total volume of the
solid is 1 cm$^3$. Helium is admitted into the sample space
through a heated filling line at the top of the TO structure. The
TO is thermally connected to a $^3$He refrigerator through its
torsion rod. Capacitive pressure gauge: the lower (moving)
capacitor plate is attached to the cell and the higher one (fixed)
is connected to a fixed support structure. }
\label{fig:TO_figure}
\end{figure}

We used several TOs in our experiments. A cross section of the
one used currently is shown in figure \ref{fig:TO_figure}. In the
current version, both the torsion rod and the annular He cell are
made of Beryllium-Copper. The loaded Q of this TO is about $5
\times 10^5$. In the earlier versions, the annular He cell was
made of 7075 Al alloy, with a Q close to $ 10^6$. The annular He
space has an outer diameter of 17mm, a width of 2mm , and a total
volume of 1cc. The TO is driven electrostatically. The tangential
rim velocity can be varied between several $\mu$ m/sec and several
cm/sec. The cell was filled with He through a heated filling line,
entering the TO structure through the top. The section of the
filling line which is part of the TO is a capillary with a volume
of around 0.001 cc, a thousand times smaller than that of the
cell. An important feature of our design is the absence of any
sharp corners inside the cell. This feature is essential in order
to enable solid He to completely fill the cell, without leaving
pockets of fluid at the corners \cite{Tuvy,BalibarAngles}. By our
estimate, filling this volume with solid He should change the
resonant period of oscillations by 0.15\%. The observed value
(4.01$\mu$sec) is in good agreement with the calculation. Our
frequency resolution allows us to determine the amount of solid He
to within 1:1000 of the full cell value. The oscillator is
thermally connected to a $^3$He refrigerator through its torsion
rod. Hence, the bottom of the cell inside the oscillator is the
coldest internal surface and the solid nucleates there.

During the measurements, we needed to measure the actual pressure
inside the cell. In our latest experiments we used a capacitive
pressure gauge shown in figure \ref{fig:TO_figure}. The moving
plate of the capacitor forming this gauge is attached to the top
of the cell, while the fixed plate of the capacitor is attached to
the support structure. As the pressure increases, the top of the
cell inflates, moving the plate upwards and increasing the
capacitance. Our gauge has a capacitance of around 15pF,  with a
noise level of several times 10$^{-5}$pF. The pressure sensitivity
of of the gauge is around 0.0289pF/bar, with a resolution better
than 1 mbar. The top of the BeCu cell, on which this gauge was
placed, had a non-hysteretic and linear pressure dependence over
our working range. In order to get the most accurate results, the
gauge was re-calibrated for each crystal grown.

In our previous TOs, the pressure inside the cell was measured in
a different way. Specifically, we monitored the resonant frequency
of a composite vibrational mode of the filling line and the
oscillator. By simulating the response of the oscillator, we
identified this vibrational mode as a combination of a "floppy"
oscillation mode of the TO (a mode where the torsion rod bends
rather than twists) with a torsional mode of the bent filling
line. This mode was observable only when the filling line was
fixed at its upper end. Reducing the length of this section of the
filling line increased the resonant frequency and lowered its
amplitude. As the pressure was
raised in the cell, the top of the cell inflated, bending the
filling line further and thus reducing its resonant frequency.

\begin{figure}
  \includegraphics{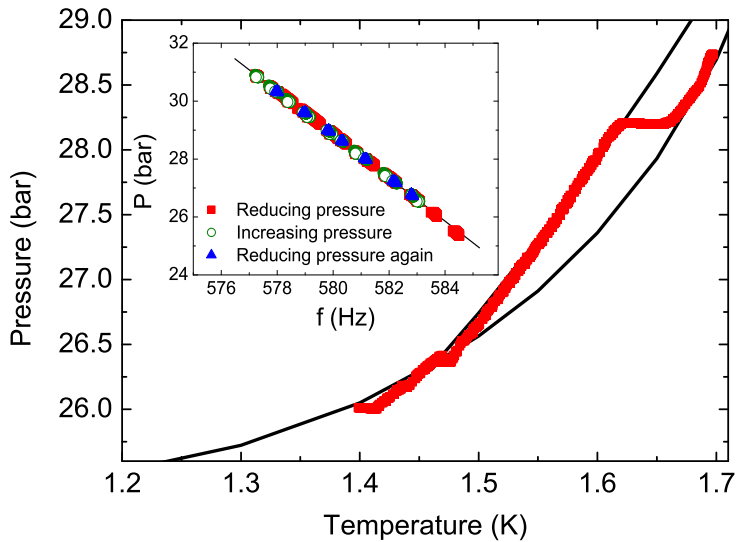}
\caption{(Color on-line) Pressure versus temperature in a cell full of solid,
measured using the pressure gauge based on the capillary vibrational mode.
The red symbols show the pressure in the cell during heating from
T=1.4K to 1.69K. This particular crystal was grown initially at 1.66K.
The solid line is the phase diagram taken from \cite{Grilly}. The
inset shows the pressure dependence of the resonant frequency of
the capillary vibrational mode. The line is a linear fit. The
red squares and empty green circles were measured at 2.5K, whereas the
blue triangles were obtained during cooldown to 1.9K. Note the
reversible behavior under pressure cycling.}
\label{fig:Aluminium_floppy545_heat_calib}
\end{figure}

This method was used with two different TO's, one made of 7075 Al
alloy, and the other from BeCu. The pressure dependence of the
resonant frequency of this mode was qualitatively the same for
both TOs. The temperature dependence of this mode was very small
and its dependence on the amount of He in the cell was negligible.
This mode was excited in the same way as the torsional mode,
however, we used a separate pair of electrodes for detection. The
moving electrode, biased at high voltage, was attached to the cell
at the insertion point of the filling line. The Q of this mode was
not as high as that of the torsional mode, and so the resonant
frequency was determined by sweeping the frequency using a signal
generator. The dependence of the resonant frequency on pressure
was calibrated while the cell was filled with liquid at a
temperature above 2K against an external pressure gauge. The
inset of figure \ref{fig:Aluminium_floppy545_heat_calib} shows an
example of such a calibration in an Aluminium cell.

The performance of this gauge is shown for example in figure
\ref{fig:Aluminium_floppy545_heat_calib}.  The figure shows the
pressure, as deduced from the resonant frequency, upon heating a
cell full of solid. For comparison, the phase diagram
\cite{Grilly} is also shown. This particular crystal was initially
grown at 1.66K and cooled to 1.4K. At 1.4K, the location on the
phase diagram was on the melting curve. We monitored the resonant
frequency as we heated the crystal. First, the solid followed the
melting line, until it reached the lower triple point. It then
continued along the bcc-hcp coexistence line, and followed it till
the cell was completely full of bcc. Then, upon heating, we
entered the bcc phase, and reached the melting line again. As can
be seen in the figure, the resolution of this pressure gauge is
entirely adequate.

This method gave reasonably good results when used in a
configuration where the filling line was made of a single piece of
capillary tubing. When the filling line was constructed of two
capillaries, one inside the other (which we needed for a different
reason), the added friction damped this
mode to the point where this method became very inaccurate.
Hence, in the current version we use a capacitive gauge
to measure pressure.


    \section{Crystal Growth} \label{sec:crystalGrowth}
The objective of our experiment was to observe the response of the
TO after introducing disorder into a single crystal. For that, we
first needed to grow single crystals of helium inside the
experimental cell. Our knowledge of single crystal growth
comes mainly from neutron scattering experiments
\cite{Oshri,Tuvy}. During these experiments elastic
scattering was used to characterize the quality and orientation
of the crystals grown. These experiments demonstrated the conditions
under which the growth results in high quality
single crystals. Similar growth conditions were used in the current
TO experiment. 

We grew crystals at constant temperature and pressure on the
melting curve. Keeping the filling line heated to several mK above
the cell temperature enabled us to work with pressures somewhat
above the melting pressure inside the cell without blocking the
capillary. To start the growth, we first raised the pressure of
the liquid inside the cell to almost that of the melting curve. We
then added a small quantity of He to the cell by opening the valve
connecting the cell to a small standard volume at room
temperature.  The initial pressure inside the standard volume was
usually around several hundred mbars above that in the cell. As He
was being added to the cell, we waited for solid nucleation to
occur. Nucleation was detected by monitoring the period of the TO,
which started to increase once solid was formed. Referring to
figure \ref{fig:TO_figure}, with the heated filling line, the
coldest spot in the cell was at its bottom, connected to the
$^3$He refrigerator through the torsion rod. We expect the crystal
to nucleate in this region. To minimize the possibility of growing
several crystals at different locations, which would result in a
poly-crystalline sample, we allowed the system to equilibrate for
one hour after the first nucleation. Optical studies on solid
$^4$He show that although very often several nuclei are created at
first, after a long enough wait only one crystallite survives, and
the rest dissolve.

Once we have waited enough to have only a single nucleus in the
cell, we commenced adding small quantities of He from the same
standard volume to the cell at constant time intervals. Each
filling cycle added a quantity of He into the cell equivalent to
about 0.03 cc of solid. After each cycle, the overpressure inside
the standard volume relaxed to the melting pressure at the growth
temperature. As long as the crystal grew the resonant period
continued to increase. Once the pressure has relaxed to the
melting pressure, the period stopped increasing (see figure 2 in
\cite{Eyal2010}). It is also possible to grow crystals
continuously, using a small (30-50 mbar) constant overpressure at
room temperature.

The absolute temperature inside the cell is determined from the
melting pressure measured in-situ using Grilly's paper
\cite{Grilly}. During crystal growth, the overpressure inside the
standard volume relaxes to the same melting pressure. This enables
us to state that within our resolution there are no pressure or
temperature gradients in our cell. When growing a crystal from the
superfluid it is relatively straightforward to achieve such
stability. When growing from the normal fluid phase above 1.772K,
the equilibration times become much longer. A typical time to grow
a single crystal from the superfluid is around 10 hours, while
with the normal fluid it can take up to 3 days. The reason is the
slow rate at which heat leaves the system in the normal fluid
phase.

There are several indications signaling that the cell is
completely full with solid. The first and most obvious one is that
the overpressure in the cell no longer relaxes to the melting
pressure. Once the cell is full, additional solid is formed in the
heated filling line, where the temperature and the melting
pressure are higher. In subsequent filling cycles, the pressure
relaxes to values which are higher and higher as the liquid in the
filling line solidifies in hotter sections. The second indication
is that the period of the TO does not change further with
subsequent filling cycles and stays constant (see figure 2 in
\cite{Eyal2010}). A third sign is the amplitude of the
oscillations, which rises systematically during the growth. During
the final stages of the growth the amplitude increases rapidly,
and eventually reaches a constant value once the cell is full.

In principle, our method of crystal growth could be used down to
the minimum of the melting curve at 0.8K. In practice, the lowest
temperature at which we could grow a crystal was constrained by
heat flow through the superfluid in the filling line. The
lowest growth temperature in this experiment was 1.07K. After
blocking the filling line the crystals could be cooled down to
temperatures around 0.5K.


\section{Disordering Procedure} \label{sec:disorderingProcedure}
Our main goal was to observe the influence of disorder on the
moment of inertia of a single crystal. To achieve that we needed
to have a controlled disordering procedure. Neutron scattering
experiments done on samples of $^4$He \cite{Oshri} in several
different cells all show that in general, the solid is very
fragile and easily disordered by application of stress. Similar
results were also recently observed in acoustic experiments
measuring the shear modulus \cite{BalibarStress,Beamish}, in which
the critical shear stress of $^4$He was found to be around
10$^{-6}$ bar. Stress of this magnitude could be generated in
several ways. One possibility which comes to mind is by cooling or by creating temperature gradients. Referring to figure \ref{fig:TO_figure}, our crystal
is grown inside an annulus around a central post. Upon cooling the
cell by $\Delta$T , the crystal should contract. However, it is
prevented from contracting by this central post, creating a strain
of $\alpha \Delta$T and stress of G $\times$ strain. Here,
$\alpha$ is the thermal expansion coefficient, and G is the shear
modulus. Putting numbers in, we find that at 1.6K it is enough to
cool the crystal by as little as 10 mK to exceed the critical
stress. Temperature gradients could be generated using the filling
line heaters. Another possibility was to lower the
temperature of the $^3$He cold stage cooling the cell. There is
also a possibility of introducing stress mechanically, by applying
a pressure pulse while the filling capillary is still open. The
results of such operations were seen in the neutron scattering
experiments. When introducing such stress to a cell completely
full of solid, a single crystal would invariably decompose into
several grains with low angle grain boundaries between them. We
believe that a similar situation occurs in our cell, with the
crystal cracking into grains which however remain oriented. An
estimate of the typical grain size is given below in the
"Results" section.

It is important to point out that as long as there is any fluid in
the cell, the procedures described above do not disorder the
crystal. In each of these scenarios, the fluid acts as a buffer
absorbing the stress, for example by solidifying during cooling
\cite{Eyal2010}.


\section{Results} \label{sec:results}
Broadly speaking, the temporal behavior of the TO after
disordering a crystal is always similar, with the resonant period
and the amplitude of oscillations both decreasing \cite{Eyal2010}. Depending on
temperature and $^3$He concentration, it can take between a few
minutes to a few hours for these changes to reach completion. Once
the period and amplitude have stabilized, their values remain
constant for as long the ambient conditions are steady. The
magnitude of the period change upon disordering the crystal can be
between several percent to about 35\% of the period change
resulting from completely filling the cell with solid. We call the
ratio between these two numbers the "mass decoupling fraction",
$\Delta I / I_{solid}$. Simultaneously, the dissipation 1/Q deduced
from the oscillation amplitude at constant drive, increases upon
breaking the crystal by a factor of 10-30.

The initial mass decoupling fraction, $\Delta I / I_{solid}$
depended on the size of the temperature or pressure shock
disordering the crystal. The crystal could be further disordered
by cooling it to a lower temperature. Ultimately, all the crystals
grown above $\sim$1.4K reached a constant maximal mass decoupling value,
$\Delta I_{max} / I_{solid}\sim$30\%-35\%, irrespective of their
crystalline symmetry. The temperature at which the crystals
reached this maximal value depended on the growth temperature and
on the initial disordering, however it seems that 35\% was
the maximal mass decoupling fraction achievable in our geometry.

\begin{figure}
  \includegraphics{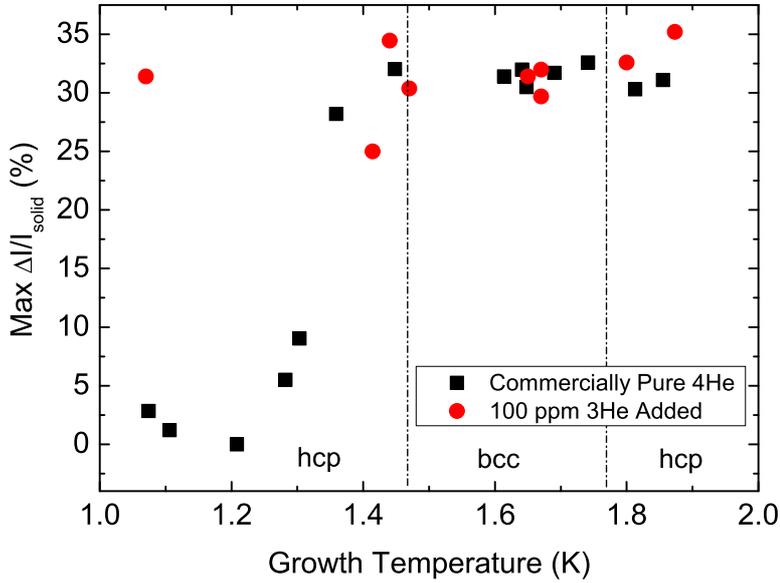}
\caption{(Color on-line) The maximal mass decoupling fraction as a function of the
growth temperature. Crystals grown from a commercially pure $^4$He
at temperatures above the lower triple point reached a final mass
decoupling fraction of around 30\% . Crystals grown below around
1.3K did not exceed 6\%. Crystals containing 100ppm $^3$He, on the
other hand, reached mass decoupling of more than 30\% even below
1.1K. } \label{fig:maximal_mass_decoupling}
\end{figure}

The only exception as far as the value of $\Delta I_{max}
/I_{solid}$ goes occurred for commercially pure $^4$He hcp crystals grown below the lower triple point of 1.464K. For these crystals, the maximal amount of
mass decoupling decreased with the growth temperature. Crystals grown below 1.3K could not reach a value higher than several percent (See figure \ref{fig:maximal_mass_decoupling}). One of these low temperature
crystals, grown at 1.21K, did not show mass decoupling at all. We
believe that the reason for that could be the orientation of the
crystal. The c facet's roughening temperature being 1.28K, implies
that at 1.21K the crystal grows faceted. If the c axis of the
crystal is not aligned with the symmetry axis of the cell, then
with a faceted crystal liquid can be trapped in a corner, and
prevent complete filling of the cell with solid. In experiments
done with the cell having sharp internal corners, where liquid was
trapped, the mass decoupling effect did not appear. That particular 1.21K sample was grown like the rest, but the
filling line was blocked when the mass load in the cell was 0.88\%
below the expected value.

\subsection{The Influence of $^3$He} \label{sec:influenceOf3He}
The low temperature TO \cite{KC2004} and shear modulus experiments \cite{Beamish} are extremely sensitive to the presence of  $^3$He. To
see to what extent $^3$He influences our results, we repeated the
TO experiment with a mixture of $^4$He and $^3$He. A priori, in
our temperature range, there is no phase separation and
crystalline defects like dislocations are not pinned by $^3$He
atoms (The binding energy of $^3$He atoms to dislocations is 0.3-0.7K \cite{Paalanen,Iwasa_binding}). We used a gas mixture with a concentration of 100ppm $^3$He. Such a low concentration of $^3$He does not change the
phase diagram appreciably. It lowers the lower triple point and
the hcp-bcc transition line by only a few mK \cite{LeeMixtures}.
However, the concentration of $^3$He is significantly higher at
surfaces and grain boundaries (GB) than in the bulk. That is because $^3$He atoms diffuse to GB's in order to lower their zero point energy, as the density at the GB's is lower than that of the bulk solid. For this reason,
the presence of even a small $^3$He concentration does
significantly lower $T_R$, the roughening temperature of the c
facet in hcp crystals \cite{EmilRoughening}. For pure $^4$He $T_R$
is 1.28K, while for crystals containing 100ppm of $^3$He $T_R$ is
around 1K.

\begin{figure}
  \includegraphics{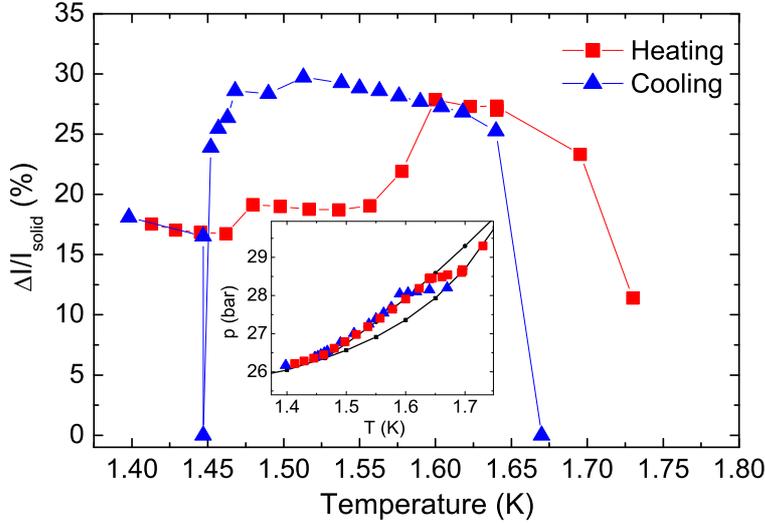}
\caption{(Color on-line) Mass decoupling fraction vs. temperature for a crystal
grown at 1.67K containing 100ppm $^3$He. The blue triangles and
red squares show mass decoupling measured during cooling and
heating respectively. During cooling, the mass decoupling vanished
at the lower triple point, while upon heating it did not. The
irreversibility in mass decoupling is pronounced along the bcc-hcp
coexistence line. There was almost no irreversibility seen with
crystals grown using commercially pure $^4$He. This irreversible
behavior can be explained by the progressive diffusion of $^3$He
from the bulk to the interfaces (GB's). The inset shows the
pressure in the cell during this process. As can be seen, within
our resolution, and aside from a small change in the melting
pressure due to creep of material into the cell at low
temperatures, the pressure in the cell is reversible.}
\label{fig:100ppm_1.67K_cool}
\end{figure}

The crystals containing 100ppm of $^3$He were grown in the same way as the
commercially pure $^4$He samples. The cell was connected through
the filling line to a bottle containing the mixture at a high
pressure. Figure \ref{fig:100ppm_1.67K_cool} shows pressure versus
temperature upon cooling a cell filled with a solid grown at
1.67K. During growth, the $^3$He concentration is uniform inside
the solid. Once the crystal has disintegrated into grains, $^3$He
starts to diffuse from the bulk onto the GB's. Here, a disintegrated crystal stands for a disordered crystal, composed of several grains misoriented by a few degrees. This diffusion is the reason for the irreversible behavior of the mass decoupling fraction between the cooling and the heating of the crystal.
During the day or so separating these processes the concentration
of $^3$He on the GB's has increased. The
irreversibility is especially evident along the bcc-hcp
coexistence line and at higher temperature where the solid reaches
the melting curve. With pure $^4$He, once the solid reaches the
melting curve the crystal anneals and the mass decoupling fraction
goes to zero. With 100ppm of $^3$He in the cell, when the solid is
heated and reaches the melting curve, a significant amount of
decoupled mass still remains, even with liquid in the cell. This
fraction would probably go to zero at some higher temperature, yet
it demonstrates that the addition of $^3$He alters some properties
of the solid which are related to the mass decoupling.

\begin{figure}
\includegraphics{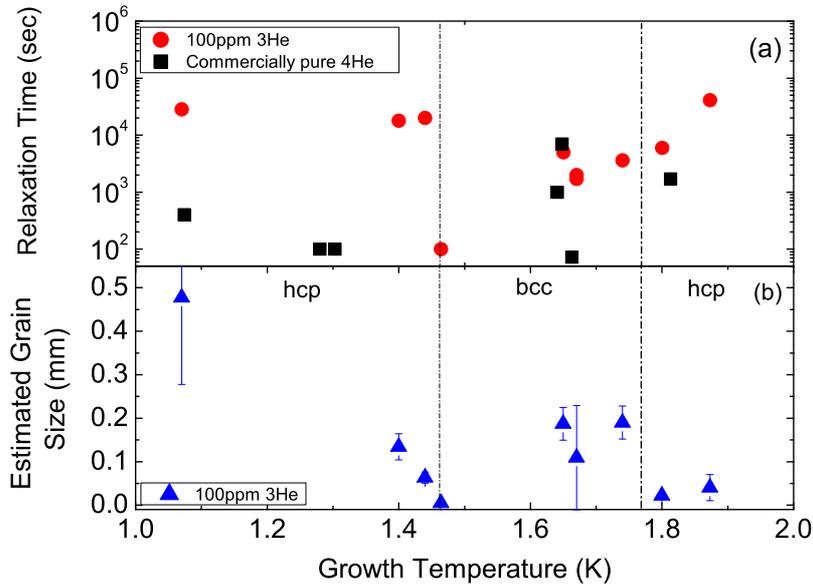}
\caption{(Color on-line) (a) Characteristic time needed for crystals to
disintegrate at their growth temperature following an application
of stress. In the bcc phase, where diffusion is fast, it took pure
$^4$He crystals (black squares) and crystals containing 100ppm
$^3$He (red circles) about the same time to relax. In the low
temperature hcp phase, however, it took the crystals containing
$^3$He more than an order of magnitude longer to relax. (b)
Typical grain size in the disordered crystal (blue triangles).
These were estimated by calculating the diffusion length over the
time scales shown in (a). } \label{fig:grain_size}
\end{figure}

The structural relaxation times for crystals containing $^3$He are equal to or larger than those measured using commercially pure $^4$He. In the bcc
phase, the relaxation times are almost the same. 
Below the lower triple point of 1.464K, crystals containing $^3$He
exhibited a different time scale for disordering. Specifically, it took more that a day (up to 2 orders of magnitude longer) to
reach a constant mass decoupling value. Figure \ref{fig:grain_size} (a) shows the time it took the crystals to reach a constant mass decoupling
value. Diffusion of $^3$He atoms
in the crystal could explain these differences. The
diffusion coefficient of $^3$He in the hcp phase is an order of
magnitude smaller than in the bcc phase \cite{Grigorev}. The longer time scale for disordering in the hcp phase is consistent with this difference. As pointed out above, $^3$He atoms diffuse from the bulk to the GB's. As as long as this mass transport takes place, structural relaxation of the crystal continues
and the mass decoupling fraction keeps increasing. By calculating the diffusion length, using
the characteristic time $\tau$ it took structural relaxation of the crystals to reach completion, we get a rough estimate of the grain size. Using values for the
diffusion coefficient $D$ \cite{Grigorev} and the relation
$<L>=\sqrt{D \cdot \tau}$ we find that the grain size $L$ is
almost always around a few parts of a mm. These results can be
seen in figure \ref{fig:grain_size} (b). The typical grain size is
consistent with the size of our cell, and with previous results
found in neutron scattering experiments \cite{Oshri}. The only
exception was a crystal grown at 1.07K, which gave a somewhat
larger grain size. This is probably due to the lower thermal
expansion coefficient at these temperatures which resulted in less
stress on the crystal upon disordering. It is interesting to point out that for a crystal grown at the lower triple point, the
relaxation time was very short (See figure \ref{fig:grain_size}
(a)). This should be the case if the crystal disintegrates into very small grains
at the triple point. Optical studies of He crystals \cite{Tuvy,Fujii} support this picture.

The addition of $^3$He also affected the maximal mass decoupling
fraction. This was only apparent below about 1.3K. Figure
\ref{fig:maximal_mass_decoupling} shows this fraction for crystals
containing both commercially pure $^4$He (black squares) and
100ppm of $^3$He (red circles) as a function of the crystal growth
temperature. As stated above, crystals grown using commercially
pure $^4$He show a drastic fall in the maximal possible mass
decoupling. In contrast, crystals containing $^3$He reached the
same maximal mass decoupling values over all our temperature
range. This could be related to the decrease of the roughening
temperature due to the presence of $^3$He \cite{EmilRoughening}.

\subsection{Velocity Dependence of the Decoupled Mass} \label{sec:velocityDependenceOfDecoupledMass}
\begin{figure}
  \includegraphics[width=1.7\textwidth]{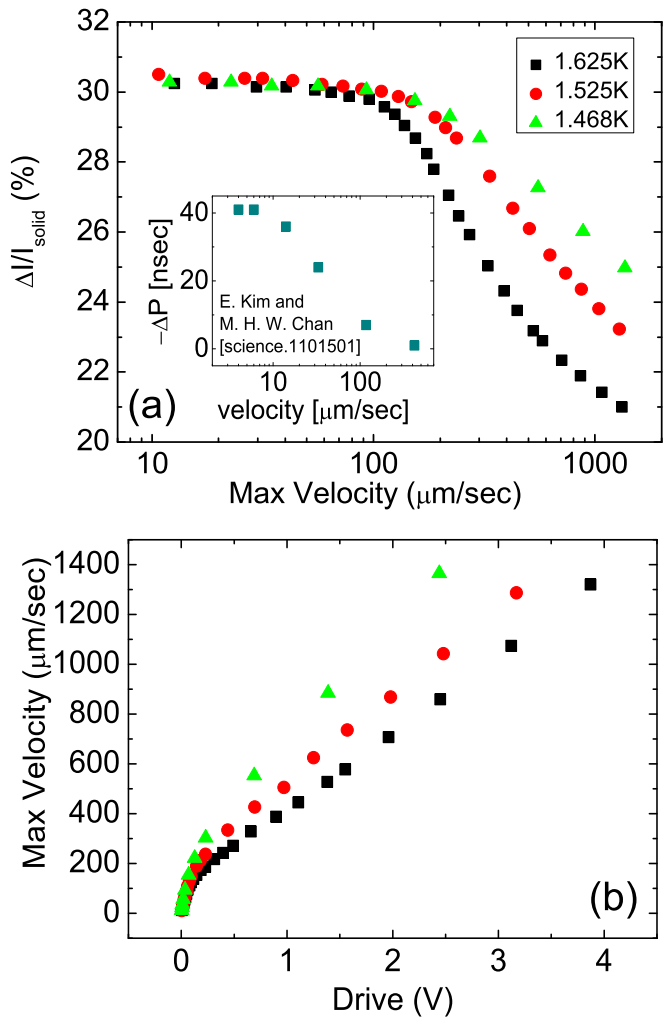}
\caption{(Color on-line) (a) Velocity dependence of the mass decoupling fraction
for a bcc crystal grown at 1.67K. Black squares - bcc crystal at
1.625K. Red circles and green triangles are data for a solid on
the bcc-hcp transition line, at 1.525K and 1.468K respectively.
For comparison, the inset shows data points of the velocity dependence
of an hcp crystal at about 0.03K taken from \cite{KCScience}. (b)
The dependence of velocity on driving force for the same data as
in (a). Two regimes are evident, the first one with higher Q and
almost no change in the mass decoupling. The second one shows
higher dissipation, and the mass decoupling is decreasing with
velocity.} \label{fig:velocity_bcc}
\end{figure}

We now describe the dependence of the mass decoupling fraction on
the rim velocity of the TO. This quantity is shown in figure
\ref{fig:velocity_bcc} (a) for a bcc crystal grown at 1.67K. At
velocities lower than $\sim$ 100 $\mu$m/sec the mass decoupling
fraction is approximately constant. At higher velocities, the mass
decoupling fraction starts to decrease. At lower temperatures,
this crossover velocity increases, and the mass decoupling
decreases with velocity at a smaller rate. For comparison, data
taken from \cite{KCScience} are shown in the inset of the figure.
As can be seen, the dependence of the period change (proportional
to mass decoupling) on the velocity is qualitatively the same,
except for the scale. Figure \ref{fig:velocity_bcc} (b) shows the
velocity as function of the driving force. The slope of the data
is inversely proportional to the dissipation of the TO. It is seen
that the lower the temperature the smaller the dissipation. The
change of the slope in Figure \ref{fig:velocity_bcc} (b) occurs at
the same crossover velocity where the mass decoupling starts to
decrease. The changes of the mass decoupling fraction with
velocity are therefore associated with the onset of increased
dissipation. Below 1K, the dissipation approaches the residual
level of the TO. This means that within our resolution, the motion
of solid He inside the TO does not add to the dissipation.

It was recently suggested that many of the phenomena seen in TO
experiments can be attributed to some properties of dislocations
in solid He \cite{Soyler,Chui,Iwasa}. In single crystals, the
density of dislocations is typically low
\cite{BeamishDislocations,BalibarStress} and so in our work we
expect to find dislocations mainly at the GB's separating adjacent
grains. Our measurements were done in a temperature regime where
$^3$He atoms do not pin dislocations, and at low dislocation
density there should be no network pinning.  We therefore expect
the dislocations in our samples to be mobile. Dislocation motion
can take place through climb or glide. For climb, vacancies are
needed. The activation energy of vacancies in bcc solid He is
around 10K. The temperature dependence of our data is inconsistent
with this activation energy. In addition, at low temperatures we
observe that the mobility of solid He is accompanied by very low
dissipation. If this mobility is mediated by dislocations, then a
conservative type of motion is needed, namely glide. Glide of
dislocations is generated by kinks which are formed by the stress
applied at the moving wall of the cell. These mechanically
generated kinks travel along the dislocation. Dissipation can take
place for example if these are scattered by thermally excited
kinks or by thermal phonons \cite{Iwasa_phonons}. Such a process can define a characteristic velocity $v^*$, such as the crossover velocity seen in figure
\ref{fig:velocity_bcc}. If one considers the kink-kink scattering only, then on dimensional grounds this characteristic velocity $v^*$ can depend on the sound velocity, the grain size, and on the density of thermal kinks, $n_k$, which
is strongly dependent on temperature (with an activation energy of
0.3K-0.4K).

\begin{equation}
v^* \propto C_s/L n_k
\end{equation}

Here $C_s$ is the speed of sound and $L$ is the typical grain
size. According to this relation, the crossover velocity above
which the dissipation of the TO increases should decrease at
higher temperatures, which is broadly consistent with experimental
results.

\section{Conclusion} \label{sec:conclusion}

It seems that the mass decoupling in the TO experiment at
temperatures between 1K and 2K has to do with structural changes
of the crystal. The dependence of the amount of decoupled mass on
the c facet's roughening temperature, as evident from the 100ppm
$^3$He experiments, indicates that surface effects are important.
Whether these effects change the mobility of grain boundaries or
the type of mobile dislocations requires further investigation.

In recent work, Reppy \cite{Reppy2010} suggests a non-supersolid
origin to the mass decoupling effect observed in TO experiments at
low temperatures. Among other issues, this work emphasizes the
importance of comparing the correct moment of inertia of the TO
with the solid in an immobile state. Our crystals show no mass
decoupling before introducing disorder, and the moment of inertia
of a cell containing a single crystal is consistent with the
classical calculation. Therefore, our experiment clearly shows
that the mass decoupling results from the mobile state and that
the decoupled mass fraction is independent of the initial state of
the solid. The connection to the low temperature experiments in
solid helium may be established if the low temperature experiments
would be done using single crystals as samples. Only then, can the
change in the period of oscillations be interpreted properly.


%

\begin{acknowledgements}
We thank L. Embon for his contribution to this work. Technical
assistance by S. Hoida, L. Yumin, and A. Post is gratefully
acknowledged. This work was supported by the Israel Science
Foundation and by the Technion Fund for Research.
\end{acknowledgements}

\bibliographystyle{spphys}       
\bibliography{supersolid}   


%
%

\end{document}